\documentclass[12pt,a4paper,superscriptaddress]{revtex4}
\usepackage{graphicx}
\usepackage{graphics}
\usepackage{amsmath}
\usepackage{dcolumn}
\usepackage{amssymb}
\usepackage{bm}
\usepackage[latin1]{inputenc}
\newcommand{\be}{\begin{equation}}
\newcommand{\ee}{\end{equation}}
\newcommand{\bea}{\begin{eqnarray}}
\newcommand{\eea}{\end{eqnarray}}

\newcommand{\pa}{\partial}

\begin{document}

\title{On the radiative corrections  in the Horava-Lifshitz $z=2$ QED}

\author{M. Gomes}
\affiliation{Instituto de F\'\i sica, Universidade de S\~ao Paulo\\
Caixa Postal 66318, 05315-970, S\~ao Paulo, SP, Brazil}
\email{mgomes,ajsilva@if.usp.br}

\author{T. Mariz}
\affiliation{Instituto de F\'\i sica, Universidade Federal de Alagoas\\ 
57072-270, Macei\'o, Alagoas, Brazil}
\email{tmariz@fis.ufal.br}

\author{J. R. Nascimento}

\affiliation{Departamento de F\'{\i}sica, Universidade Federal da 
Para\'{\i}ba\\
 Caixa Postal 5008, 58051-970, Jo\~ao Pessoa, Para\'{\i}ba, Brazil}
\email{jroberto,petrov@fisica.ufpb.br}

\author{A. Yu. Petrov}

\affiliation{Departamento de F\'{\i}sica, Universidade Federal da 
Para\'{\i}ba\\
 Caixa Postal 5008, 58051-970, Jo\~ao Pessoa, Para\'{\i}ba, Brazil}
\email{jroberto,petrov@fisica.ufpb.br}

\author{A. J. da Silva}
\affiliation{Instituto de F\'\i sica, Universidade de S\~ao Paulo\\
Caixa Postal 66318, 05315-970, S\~ao Paulo, SP, Brazil}
\email{mgomes,ajsilva@if.usp.br}

\begin{abstract}
We calculate  one-loop contributions to the two and  three point spinor-vector functions in $z=2$ Horava-Lifshitz QED. This allows us to obtain the anomalous magnetic moment.
\end{abstract}

\maketitle

\section{Introduction}

Studies of Lorentz symmetry breaking, together with searches for renormalizable gravity models, have aroused interest on field theory models with space-time anisotropy.
The paradigmatic example of such  theories is the Horava-Lifshitz (HL) gravity \cite{Hor} characterized by the fact that, while the action continues to be of the second order in time derivatives, which is necessary to avoid the occurrence of  ghost states, higher orders in spatial derivatives are used. As a result, the convergence of quantum corrections is  improved what gives  hope for the possibility to construct a renormalizable ghost-free gravity theory. 

Besides the interest on gravity, the HL-like extensions to other field theory models are presently being intensively studied. It is worth to mention that, originally the space-time asymmetry emerged within studies in statistical mechanics \cite{Lif} which clearly motivates  further its application to condensed matter \cite{Fradkin} and other contexts. The most important results achieved in these studies are the proof of renormalizability of HL-like scalar field models \cite{Anselmi}, explicit calculation of counterterms in HL-like QED \cite{ed} and other theories \cite{gomes}, calculation of the effective potential in different HL-like scalar theories, and in Yukawa-like theory and QED, including the finite temperature case \cite{ff,Farias} (for a review on HL-like field theory models, see also \cite{PRSG}). 
Furthermore, in our previous paper \cite{prev}, we provided an analysis of Lorentz symmetry restoration and the existence of anomalies in $z=2$ spinor and scalar QED. A natural continuation of this study could consist in obtaining the HL-like analogue of the anomalous magnetic moment. This is the problem addressed in the present paper.

The structure of this work is as follows. In the section 2, we describe the classical action, propagators and vertices of $z=2$ QED. In the section 3 we calculate the contributions to the three-point vector-spinor function and to the quadratic part of the fermionic Lagrangian. Based on these results, the one-loop renormalization of the model is analyzed. A Summary is devoted to the discussion of the results.

\section{Classical action, propagators and vertices}

We consider the Horava-Lifshitz-like (HL-like) spinor QED.
For the sake of  concreteness, we restrict ourselves to the case $z=2$, as in \cite{prev}. 
In this case, the Lagrangian  describing  the model we are interested is
\bea
\label{scal}
L&=&\frac{1}{2}F_{0i}F_{0i}+\frac{b^{2}}{4}F_{ij}\Delta F_{ij}+\bar{\psi}(i\gamma^0D_0+a(i\gamma^iD_i)^2-m^2)\psi,
\eea
where $D_{0,i}=\pa_{0,i}-ieA_{0,i}$ is a gauge covariant derivative, with the corresponding gauge transformations being  $\psi\to e^{ie\xi}\psi$, $\bar{\psi}\to \bar{\psi}e^{-ie\xi}$, and $A_{0,i}\to A_{0,i}+\pa_{0,i}\xi$. Our metric is $(+,-,-, -)$, and $\Delta$ denotes the $d$-dimensional Laplacian. 
The parameters $a$ and $b$ were introduced to keep track of the contributions associated to the higher derivative terms.   

The free propagator of the fermionic field is
\bea
\label{freeprop}
G(k)=<\psi(k)\bar{\psi}(-k)>&=&i\frac{\gamma^0k_0+\omega}{k^2_0-\omega^2+i\epsilon}\nonumber\\
&=&\frac{P_+}{k_0-\omega+i\epsilon}-\frac{P_-}{k_0+\omega-i\epsilon},
\eea
where $\omega=a\vec{k}^2+m^2$, and $P_{\pm}=\frac{1\pm\gamma_0}{2}$ are orthogonal projectors.
To find the propagator for the vector field, we choose to work in  the Feynman gauge by adding to (\ref{scal})  the gauge fixing Lagrangian \cite{Farias},
\bea
{\cal L}_{gf} = -\frac12[ (-b^{2}\triangle)^{-\frac12}\partial_{0}A_{0}-(-b^{2}\triangle)^{\frac12}\partial_{i}A_{i}]^2,
\eea
yielding to the free propagators  the  forms
\bea 
\Delta_{ij}(k)=<A_iA_j>=\frac{i\delta_{ij}}{k^2_0-b^{ 2}\vec{k}^4+i\epsilon}; \quad\, \Delta_{0}(k)=<A_0A_0>=i\frac{b^{2}\vec{k}^2}{k^2_0-b^2\vec{k}^4+i\epsilon}.
\eea

The interaction vertices are
\bea
\label{vertspin}
V_1&=&e\bar{\psi}\gamma^0\psi A_0,\quad\, V_2=-e^2a\bar{\psi}\psi A_iA_i,\nonumber\\
V_3&=&-ieaA_i(\bar{\psi}\pa_i\psi-\partial_i\bar{\psi}\psi),\quad\, V_4=\frac{ea}{2}\bar{\psi}\sigma^{ij}\psi F_{ij},\label{8}
\eea
where  $\sigma^{ij}=\frac{i}{2}[\gamma^i,\gamma^j]$. In the momentum space they look like
\bea
V_1&=&e\bar{\psi}(p_2)\gamma^0\psi(p_1) A_0(-p_1-p_2),\quad\, V_2=-e^2a\bar{\psi}(p_2)\psi(p_1)A_i(k)A_i(-k-p_1-p_2); \label{9}\\
V_3&=&ea(p_2-p_1)_i\bar{\psi}(p_2)\psi(p_1) A_i(-p_1-p_2),\quad\, V_4=iea(p_1+p_2)_i\bar{\psi}(p_2)\sigma^{ij}\psi(p_1)A_j(-p_1-p_2),\nonumber
\eea
 with all momenta  chosen to be entering the corresponding vertices.

The mass dimensions for the case $d=3$ on which we concentrate in this paper are: $1/2$ for $e$, $3/2$ for $A_0$, $1/2$ for $A_i$ and $3/2$ for $\psi$ (as usual for HL theories, we have $1$ for $\partial_i$ and $z=2$ for $\partial_0$). Thus, we can conclude that unlike the five-dimensional theory considered in \cite{Iengo}, our theory is super-renormalizable. Indeed,  the  degree of superficial divergence for a generic graph in $d$ space dimensions can be shown to be
\bea
\omega=d+2-(\frac{d-2}{2})E_i-\frac{d}{2}(E_0+E_{\psi})+(\frac{d}{2}-2)(V_1+2V_2+V_3+V_4),
\eea
where $E_0,E_i,E_{\psi}$ are the numbers of external $A_0,A_i,\psi$ legs respectively, $V_{1,2,3,4}$ are the numbers of corresponding vertices. Thus, QED with $z=2$ is super-renormalizable if  $d<4$, renormalizable if $d=4$ and non-renormalizable if $d>4$. Various interesting aspects as renormalization and infrared properties of the theory with $d=4$ were treated in \cite{Iengo}. Motivated by the
physical relevance of the usual ($z=1$) QED in $d=3$ space dimensions, here  we will consider its extension to $z=2$. In particular, we will
focus on the anomalous magnetic moment of the fermionic field. Needless to say, our interest in this issue comes from the fact that the computation of the electron magnetic moment in relativistic QED is one of the hallmarks in  the development of field theory. In our case, the calculation is more complex because  the additional derivative couplings  lead us to proceed a complete analysis of the one loop corrections of all trilinear  vertexes. However, as our model is  super-renormalizable these corrections are finite.

From the tree approximation to the vertex $\bar{\psi}(p_2)\sigma^{ij}\psi(p_1) F_{ij}(p)$, where $F_{ij}$  is an external magnetic field, we may extract the normal magnetic moment as being $2ea \bar{\psi}\vec{S}\psi$, where $\vec{S}$ is the spin operator.

\section{Anomalous magnetic moment in the spinor QED}

To proceed with the one-loop analysis we will now study low momentum corrections to the  vector-spinor three point function and in particular determine its contribution to  the anomalous magnetic moment. That study is based  in the Feynman diagrams depicted in the Figs. 1 and 2.

The twelve possible contributions to the effective Lagrangian coming from Fig. 1 are:
\bea
\label{t-all}
T_1&=&-e^3\int\frac{d^dkdk_0}{(2\pi)^{d+1}}\bar{\psi}(p_2)\gamma^0G(k)
\gamma^0A_0(p)G(k+p)\gamma^0\psi(p_1)\Delta_{0}(k-p_{2});\nonumber\\
T_2&=&iae^3\int\frac{d^dkdk_0}{(2\pi)^{d+1}}\bar{\psi}(p_2)\gamma^0G(k)
p_iA_j(p)\sigma^{ij}G(k+p)\gamma^0\psi(p_1)
\Delta_{0}(k-p_{2});\nonumber\\
T_3&=&-a^2e^3\int\frac{d^dkdk_0}{(2\pi)^{d+1}}\bar{\psi}(p_2)G(k)\gamma^0
A_0(p)G(k+p)\psi(p_1)
(k_i+p_{2i})\nonumber\\&&\times(k_k+p_k-p_{1k})\Delta_{ik}(k-p_{2});\nonumber\\
T_4&=&ia^2e^3\int\frac{d^dkdk_0}{(2\pi)^{d+1}}\bar{\psi}(p_2)G(k)
\gamma^0A_0(p)G(k+p)\sigma^{ij}(-k_i+p_{2i})\nonumber\\&&\times\Big[(k_k+p_k-p_{1k})\Delta_{jk}(k-p_2)-(k_{k}+p_{2k})\Delta_{kj}(k-p_{2})\Big]\psi(p_1);\nonumber\\
T_5&=&a^3e^3\int\frac{d^dkdk_0}{(2\pi)^{d+1}}\bar{\psi}(p_2)G(k)
p_m\sigma^{mn}A_n(p)G(k+p)\sigma^{ij}(k_i-p_{2i})\nonumber\\&&\times\Big[(k_k+p_k-p_{1k})\Delta_{jk}(k-p_{2})-(k_{k}+p_{2k})\Delta_{kj}(k-p_{2)}\Big]\psi(p_1);\nonumber\\
T_6&=&ia^3e^3\int\frac{d^dkdk_0}{(2\pi)^{d+1}}\bar{\psi}(p_2)G(k)p_l\sigma^{lj}A_j(p)G(k+p)\psi(p_1)\nonumber\\&&\times
(k_i+p_{2i})(k_k+p_k-p_{1k})\Delta_{ik}(k-p_{2});\nonumber\\
T_7&=&ia^3e^3\int\frac{d^dkdk_0}{(2\pi)^{d+1}}\bar{\psi}(p_2)G(k)
(2k_i+p_i)A_i(p)G(k+p)\sigma^{lj}(k_l-p_{2l})\nonumber\\&&\times\Big[(k_k+p_k-p_{1k})\Delta_{kj}(k-p_{2})-(k_{k}+p_{2k})\Delta_{kj}(k-p_2)\Big]\psi(p_1);\nonumber\\
T_8&=&-a^3e^3\int\frac{d^dkdk_0}{(2\pi)^{d+1}}\bar{\psi}(p_2)G(k)
(2k_l+p_l)A_l(p)G(k+p)\psi(p_1)\nonumber\\&&\times
(k_i+p_{2i})(k_k+p_k-p_{1k})\Delta_{ik}(k-p_{2});\nonumber\\
T_9&=&-e^3a\int\frac{d^dkdk_0}{(2\pi)^{d+1}}\bar{\psi}(p_2)\gamma^0G(k)(2k_i+p_i)A_i(p)\nonumber\\&&\times
G(k+p)\gamma^0 \Delta_0(k-p_{2})
\psi(p_1);
\label{s456}
\eea
\bea
T_{10}&=&e^3a^2\int\frac{d^dkdk_0}{(2\pi)^{d+1}}\bar{\psi}(p_2)\sigma^{ij}G(k)
\gamma^0A_0(p)G(k+p)\sigma^{lk}\psi(p_1)\nonumber\\&&\times(k-p_2)_i(k-p_2)_k \Delta_{jl}(k-p_{2});\nonumber\\
T_{11}&=&-ie^3a^3\int\frac{d^dkdk_0}{(2\pi)^{d+1}}\bar{\psi}(p_2)\sigma^{ij}G(k) 
p_m\sigma^{mn}A_n(p)G(k+p)\sigma^{lk}\psi(p_1)\nonumber\\&&\times(k-p_2)_i(k-p_2)_k\Delta_{jl}(k-p_{2});\nonumber\\
T_{12}&=&e^3a^3\int\frac{d^dkdk_0}{(2\pi)^{d+1}}\bar{\psi}(p_2)\sigma^{ij}G(k)
(2k_m+p_m)A_m(p)G(k+p)\nonumber\\&&\times\sigma^{lk}\psi(p_1)(k-p_2)_i(k-p_2)_k\Delta_{jl}(k-p_{2}).\nonumber
\eea
Here  $p=-(p_1+p_2)$ is the momentum entering with the external gauge field. As indicated, the spatial parts of these integrals are dimensionally regularized  with the  parameter  $d$ attaining  its physical value, $d=3$, at the end of the calculation.

The contributions from Fig. 2  look like
\bea
T_{13}&=&-2i e^3a^2\int\frac{d^dkdk_0}{(2\pi)^{d+1}}\bar{\psi}(p_2)
A_i(p)G(k)
\psi(p_{1})\nonumber\\&&\times\Big[(k_j-p_{1j})\Delta_{ij}(k+p_{1})-(k_j-p_{2j})\Delta_{ij}(k+p_{2})\Big];\nonumber\\
T_{14}&=& 2e^3a^2\int\frac{d^dkdk_0}{(2\pi)^{d+1}}\Big[\bar{\psi}(p_2)
A^l(p)\sigma^{ij}k_i\Delta_{lj}(k)
\nonumber\\&&\times \Big[G(-k-p_1)- G(p_{2}-k)\Big]\psi(p_1).
\eea

Up to first order in the momenta, we may summarize the result of the calculations of the  previous expressions  as $T_{i}=\bar \psi {\cal T}_{i}\psi$, $i=1,2,\ldots 14$, with $s=a+b$, so,

\begin{table}[ht]
\centering
\begin{tabular*}{700pt}{l l l}
${\cal T}_{1}= \frac{e^{3}b}{16s^{3/2}m\pi}\gamma_{0} A^{0},$ &  ${\cal T}_{2}=\frac{e^{3}ab}{32s^{3/2}m\pi}\sigma_{ij} F^{ij} $&$ {\cal T}_{3}=\frac{e^{3}a^{2}}{16bs^{3/2}m\pi}\gamma_{0} A^{0},$\\[10pt]
${\cal T}_{6}= \frac{e^{3}a^{3}}{32b s^{3/2}m\pi}\sigma_{ij} F^{ij}$, & ${\cal T}_{7}=\frac{e^{3}a^{3}}{24bs^{3/2}m\pi}\sigma_{ij} F^{ij},$ &$ {\cal T}_{8} = \frac{ e^3 a^{3}(4a+7b)}{48bs^{5/2}m\pi}(p_2-p_1)_{l}A_{l}$,\\[10pt]
${\cal T}_{9}= \frac{ e^3a b^{2} }{16 s^{5/2}m\pi}(p_2-p_1)_{l} A_{l}, $ & ${\cal T}_{10}=\frac{e^{3}a^{2}}{8bs^{3/2}m\pi}\gamma_{0} A^{0},$
 & ${\cal T}_{11}=-\frac{e^{3}a^{3}}{48bs^{3/2}m\pi}\sigma_{ij} F^{ij},$\\[10pt]
 ${\cal T}_{12} =\frac{e^3 a^{3} }{8s^{5/2}m\pi}(p_2-p_1)_{l}A_{l},$  & ${\cal T}_{13}= -\frac{ e^3a^{2}(5a+6b)}{12bs^{3/2} m\pi}(p_{2}-p_{1})_{l}A_{l},$ & ${\cal T}_{14}=-\frac{e^{3}a^{3}}{24bs^{3/2}m\pi}\sigma_{ij} F^{ij}.$ 
\end{tabular*}
\end{table}
\noindent
Notice that, up to the order that we considered, $T_{4}$ and $T_{5}$ vanish (this is so because, at zeroth order in the momenta, they involve the product $\sigma^{ij}\delta_{ij}=0$).
Let us then analyze the remaining contributions.   First of all,  $T_1,T_3,T_{10}$ involve only $A_0$ and cannot yield the anomalous magnetic moment. Instead, they produce 
\begin{equation}
T_{1}+T_{3}+T_{10}= \frac{e^{3}(b^{2}+3a^{2})}{16bs^{3/2}m\pi}\bar{\psi}\gamma_{0}\psi A^{0}.
\end{equation}
 
 Also, $T_8,T_9,T_{12}$ and $T_{13}$ give
 \begin{equation}
T_{8}+T_{9}+T_{12}+T_{13}=-\frac{ e^3 a (16 a^3 + 31 a^2 b +24ab^2-3 b^3)  }{48 b s^{5/2}m\pi}(p_2-p_1)_{l}\bar{\psi}\psi A_{l}.
 \end{equation} 

Finally, we found the following contribution for the vertex $V_{4}$:
\begin{equation}
T_2+T_6+T_{7}+T_{11}+T_{14}=\frac{e^{3}a(a^{2}+3b^2)}{96 bs^{3/2}m\pi}\bar{\psi}\sigma_{ij}\psi F^{ij}.
\end{equation}
 
 Thus, up to one-loop order, the complete triple interaction term being the sum of $V_{1},\, V_{3}$ and $V_{4}$ is modified to
\begin{eqnarray}
{\cal L}_3&=&  e( 1+ {e^2}F)\bar{\psi}\gamma^0 \psi A_{0},-ie a( 1-e^2 H )A_i(\bar{\psi}\pa_i\psi -\partial_i\bar{\psi}\psi)+\\  
&+&\frac{ea}{2}( 1+ {e^2}G)\bar{\psi}\sigma_{ij}\psi F^{ij},\nonumber
\end{eqnarray}
where $F=\frac{(b^{2}+3a^{2})}{16bs^{3/2}m\pi}$,  $G=\frac{( a^2+3b^2)}{96bs^{3/2}m\pi}$ and $H=(16 a^3 + 31 a^2 b + 24 a b^2 - 3 b^3) /(48 b s^{5/2} m \pi)$.

As argued in \cite{prev}, the gauge field $A_{0,i}$ is not renormalized, and $b$ does not have radiative corrections.
For the fermion two-point function, the relevant graphs are shown at Fig. 4, but within the dimension regularization only the graph with two vertices has a non-vanishing contribution. The one-loop corrected Lagrangian of the spinor looks like
\begin{equation}
\label{eq16}
{\cal L}_{0\psi}= \bar{\psi}\Big[(1+ e^2 F)i\gamma^{0}\partial_{0}+a (1-e^2H) \triangle-m^2 (1+2e^2 F)\Big] \psi. 
\end{equation}
Notice that these results are consistent with the gauge invariance of the model. In fact, for $z=2$ there are two independent gauge invariant interactions, namely, $a \bar{\psi}(i\gamma^iD_i)^2\psi$, that we have considered, and $a_1\bar{\psi}(D_i)^2\psi$. Observe that the inclusion of the last interaction has the effect of changing $a$ to $a+ a_1$ in the vertices  $V_{2}$ and $V_{3}$ and in the  quadratic part of the fermionic Lagrangian. However, it has no effect in the vertices $V_{1}$ and $V_{4}$. For simplicity, we do not consider the $a_1\bar{\psi}(D_i)^2\psi$ vertex for the moment. We note that although we did not calculate explicitly the correction to the vertex $V_2$, the gauge invariance requires the same renormalization as in the vertex $V_3$ and in the Laplacian term occurring in (\ref{eq16}).

The renormalization of the model follows by making a reparametrization  as follows
\begin{eqnarray}
&&\psi \quad \rightarrow\quad Z^{1/2} \psi,\qquad \bar\psi \quad \rightarrow\quad Z^{1/2} \bar\psi,\qquad e\rightarrow \frac{Z_{e}}{Z}e,\nonumber\\
&& a\quad \rightarrow \quad \frac{Z_a}{Z}a,\quad\qquad m^2 \quad\rightarrow\quad m^2 + \delta m^2.
\end{eqnarray}
Therefore, written in terms of the renormalized quantities, the fermionic Lagrangian is
\begin{eqnarray}
 {\cal L}&=& Z\bar{\psi}\gamma^0 \partial_0 \psi+a Z_{a}\bar{\psi}\triangle\psi + e Z_{e}\bar{\psi}\gamma^{0}\psi A_{0}+\frac{ea}{2}\frac{Z_{a}Z_{e}}{Z}\bar{\psi}\sigma^{ij}\psi F_{ij}- i e a\frac{Z_{a}Z_{e}}{Z} A_{i}(\bar{\psi}\partial_{i}\psi-\partial_{i}\bar{\psi}\psi)\nonumber\\
 && -e^{2}a\frac{Z_{a}Z_{e}}{Z}\bar{\psi}\psi A_{i}A_{i}-(m^2+\delta m^2)Z\bar{\psi}\psi,
 \end{eqnarray} 
where the renormalization constants are chosen so that the two point vertex function satisfies
\begin{equation}
 \Gamma^{(2)}_{(p0=0,\vec p=0)}=-i m^2,\qquad \frac{\partial\Gamma^{(2)}}{\partial p_{0}}\Big\vert_{(p0=0,\vec p=0)}=i \gamma_0,\qquad \frac{\partial^{2}\Gamma^{(2)}}{\partial \vec p^{2}}\Big\vert_{(p0=0,\vec p=0)}= i a,
\end{equation}
and also the three point vertex function (one  $A_0$ field and a pair $\bar \psi$, $\psi$)   at zero momenta obeys
$\Gamma^{(2,1)}=i e$. Using these conditions, we get
\begin{equation}
Z=Z_{e}=1 - e^{2} F,\qquad Z_{a}=1+ e^2 H, \qquad \delta m^2=- e^2 m^2 F.\label{eq19}
\end{equation}

Summarizing, the renormalized Lagrangian (up to one-loop) is given by:
\begin{eqnarray} 
 {\cal L}&=& \frac{1}{2}F_{0i}F_{0i}+\frac{b^{2}}{4}F_{ij}\Delta F_{ij}+\bar{\psi}(i\gamma^0 \partial_0+a \triangle{\psi} 
-m^2){\psi}+{\cal L}_{gf}+ e \bar{\psi}\gamma^{0}\psi A_{0}\nonumber\\
&&- i e a A_{i}(\bar{\psi}\partial_{i}\psi-\partial_{i}\bar{\psi}\psi)
  +\frac{ea}{2}\bar{\psi}\sigma^{ij}\psi F_{ij} (1+e^2 H+e^2 G)\nonumber\\
&&-e^{2}a\bar{\psi}\psi A_{i}A_{i}(Z_{a}+\mbox{one$-$loop\ corrections})+(\mathrm{higher\, derivatives}).
\end{eqnarray} 

The anomalous magnetic moment induced by the radiative corrections can be read immediately from this Lagrangian.
It must be noted that no correction of the form $i\bar{\psi} \gamma_k \partial_k\psi$ or $\frac{1}{4} F_{ik}F_{ik}$ was induced,  what makes  Lorentz symmetry restoration to be a very improbable goal (this is in agreement with the results of \cite{ed} for pure boson field models and for the fermionic QED in 1+4 dimensions). However, Lorentz symmetry could possibly emerge at low energies  if the model included from the beginning the term $i\bar{\psi} \gamma_k \partial_k\psi$.

After this renormalization procedure the magnetic vertex $V_{3}$ becomes
\begin{equation}
\frac{ea}{2}(1+e^2H+e^2G) \bar{\psi}\sigma_{ij}\psi F^{ij}.
\end{equation}
This is our final result for the  magnetic moment up to one-loop.

\section{Summary}

Motivated by the fundamental role played by anomalous magnetic moment in the development of relativistic ($z=1)$ QED, here   we calculated the $z=2$ analogous term.  Due to the existence of various new vertices,
the calculation turns out to be more involved than in the $z=1$ QED, what prompted us to proceed a complete one loop analysis of the renormalizarion process.    Our computation shows an essential difference with the one in the usual QED -- while in the usual QED the one-loop $<\bar{\psi}A\psi>$ triangle diagram contributes not only to the magnetic moment but also to the renormalization of the electric charge, in this case this contribution yields no divergences, being finite from the very beginning, and even finite renormalization of the electric charge does not occur at the one-loop order.  This is a consequence of the fact that the Ward identity $Z=Z_e$ is obeyed (see \ref{eq19}) and that in this model the gauge field is not renormalized, which, in its turn, implies that the Lorentz symmetry restoration cannot occur. Thus, our study   may be  relevant at very high energies  where Lorentz invariance is presumably broken. On the other hand, at low energies the inclusion of the term linear in the spatial derivatives is  certainly mandatory.

A natural continuation of this study could consist in obtaining the possible contributions to the anomalous magnetic moment in HL-like QED with higher values of $z$, as well as in higher spatial dimensions where the theory will by renormalizable rather than super-renormalizable. We plan to carry out this study in a forthcoming paper.

{\bf Acknowledgements.} This work was partially supported by Conselho
Nacional de Desenvolvimento Cient\'{\i}fico e Tecnol\'{o}gico (CNPq)
and Funda\c{c}\~{a}o de Amparo \`{a} Pesquisa do Estado de S\~{a}o
Paulo (FAPESP). The work by A. Yu. P. has been partially supported by the
CNPq project No. 303783/2015-0.

\begin{figure}[!ht]
\begin{center}
\includegraphics[angle=0,scale=1.00]{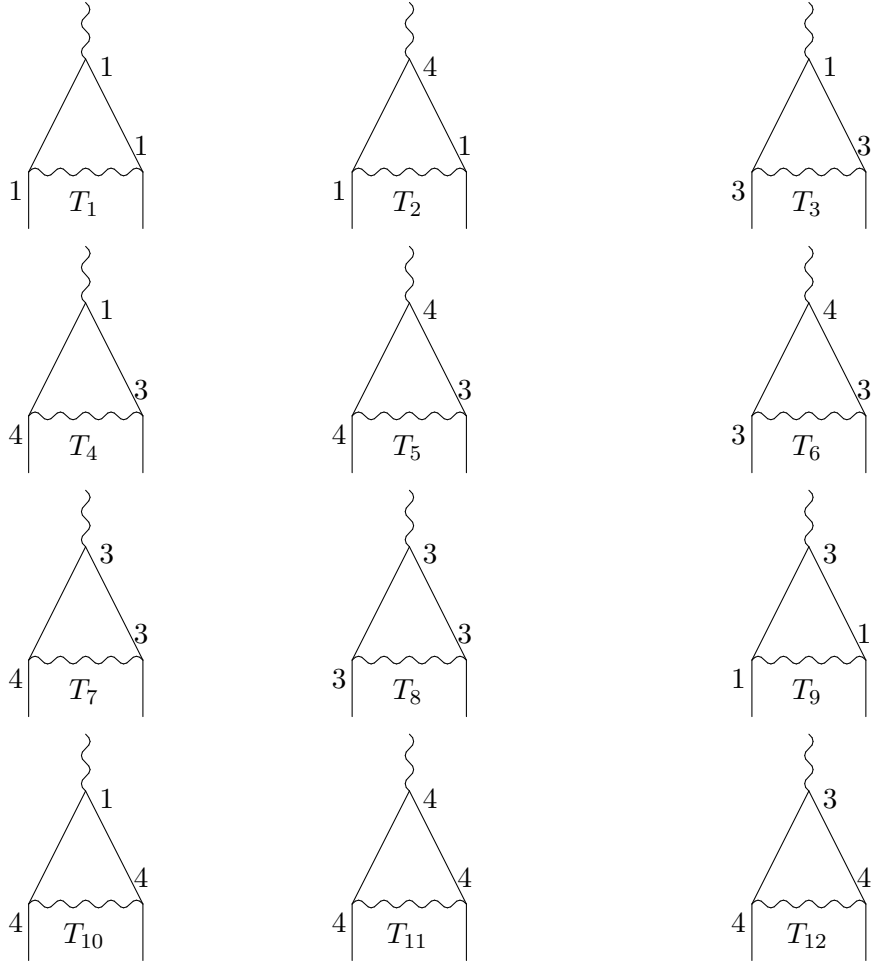}
\end{center}
\caption{One-loop, three-vertex diagrams contributing to the three point function. Here $T_i$, $i=1\ldots 12$ denote just the contributions from the expression (\ref{s456}), and the numbers 1,3,4 are for vertices $V_1,V_3,V_4$. Notice that besides these diagrams there are other graphs not shown  obtained by exchanging the vertices $V_3$ and $V_4$ in $T_4,\,T_5$ and $T_7$.}
\end{figure}

\vspace*{3mm}

\begin{figure}[!ht]
\begin{center}
\includegraphics[angle=0,scale=1.00]{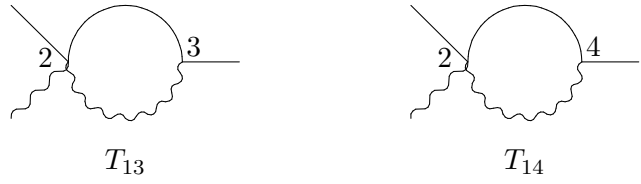}
\end{center}
\caption{One-loop diagrams with one  quartic and one triple vertices.}
\end{figure}

\vspace*{3mm}

\begin{figure}[!ht]
\begin{center}
\includegraphics[angle=0,scale=1.00]{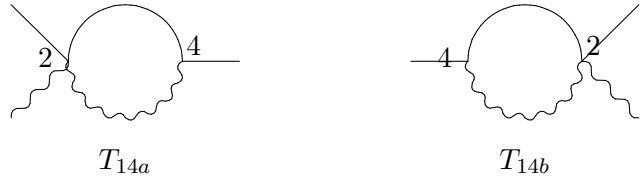}
\end{center}
\caption{Two possible contributions to $T_{14}$.}
\end{figure}

\begin{figure}[!ht]
\begin{center}
\includegraphics[angle=0,scale=1.00]{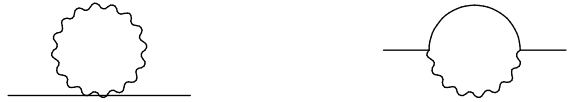}
\end{center}
\caption{One-loop diagrams contributing to the fermion two point function.}
\end{figure}
\end{document}